\documentclass{article}
\usepackage{spconf,amsmath,graphicx}
\usepackage{multirow}
\usepackage{colortbl}
\usepackage{url}
\usepackage{amssymb}
\usepackage{enumitem}
\setlist{nosep, leftmargin=14pt}
\title{Fourier Disentangled Multimodal Prior Knowledge Fusion for Red Nucleus Segmentation in Brain MRI}

\name{
\parbox{0.9\linewidth}{
\centering
Guanghui Fu $^{\dagger}$ \qquad 
Gabriel Jimenez $^{\dagger}$ \qquad 
Sophie Loizillon $^{\dagger}$ \qquad 
Rosana El Jurdi $^{\dagger}$ \qquad 
Lydia Chougar $^{\dagger\star}$  \qquad
Didier Dormont $^{\dagger\star}$ \qquad
Romain Valabregue $^{\ddagger}$ \qquad 
Ninon Burgos $^{\dagger}$ \qquad 
Stéphane Lehéricy $^{\ddagger\star}$ \qquad
Daniel Racoceanu $^{\dagger}$ \qquad
Olivier Colliot $^{\dagger}$ \qquad
the ICEBERG Study Group$^{\ddagger}$}}

\address{$^{\dagger}$ Sorbonne Université, Institut du Cerveau - Paris Brain Institute - ICM, CNRS, Inria, Inserm, AP-HP, \\ Hôpital de la Pitié Salpêtrière,  F-75013, Paris, France \\
$^{\ddagger}$ Sorbonne Université, Institut du Cerveau - Paris Brain Institute - ICM, CNRS, Inserm, AP-HP, \\ Hôpital de la Pitié Salpêtrière, F-75013, Paris, France \\
$^{\star}$ AP-HP, Hôpital de la Pitié Salpêtrière, DMU DIAMENT, Department of Neuroradiology, \\ F-75013, Paris, France}

\begin{document}

\maketitle
\begin{abstract}
Early and accurate diagnosis of parkinsonian syndromes is critical to provide appropriate care to patients and for inclusion in therapeutic trials.
The red nucleus is a structure of the midbrain that plays an important role in these disorders. It can be visualized using iron-sensitive magnetic resonance imaging (MRI) sequences. 
Different iron-sensitive contrasts can be produced with MRI. Combining such multimodal data has the potential to improve segmentation of the red nucleus.
Current multimodal segmentation algorithms are computationally consuming, cannot deal with missing modalities and need annotations for all modalities. 
In this paper, we propose a new model that integrates prior knowledge from different contrasts for red nucleus segmentation. 
The method consists of three main stages. First, it disentangles the image into high-level information representing the brain structure, and low-frequency information representing the contrast.   
The high-frequency information is then fed into a network to learn anatomical features, while the list of multimodal low-frequency information is processed by another module.
Finally, feature fusion is performed to complete the segmentation task.
The proposed method was used with several iron-sensitive contrasts (iMag, QSM, R2*, SWI).
Experiments demonstrate that our proposed model substantially outperforms a baseline UNet model when the training set size is very small. 
\end{abstract}


\section{Introduction} \label{sec:intro}

Parkinsonism is clinically defined by the association of bradykinesia, plastic rigidity and/or asymmetrical resting tremor. Parkinson’s disease is the most common neurodegenerative cause of parkinsonism. Atypical parkinsonism includes tauopathies (progressive supranuclear palsy [PSP] and corticobasal degeneration) and synucleinopathies (multiple system atrophy [MSA] and dementia with Lewy body [DLB]) \cite{dickson2012parkinson}. Early and accurate diagnosis of parkinsonian syndromes is critical to provide appropriate care to patients and for inclusion in therapeutic trials. The red nucleus is a small nucleus located in the midbrain. It is a region of particular interest considering its known pathophysiological changes in PSP \cite{williams2009progressive}. Iron-sensitive MRI can be used to assess iron deposition in the brain, particularly in the red nucleus. It provides promising biomarkers for differentiating parkinsonism and for longitudinal monitoring of disease progression and treatment effects \cite{sjostrom2017quantitative}. Different iron-sensitive contrasts can be derived from multi-echo T2* acquisitions by combining the phase and magnitude information. These contrasts include quantitative susceptibility mapping (QSM), R2* relaxation rate mapping (R2*), magnitude images (iMag) and susceptibility-weighted images (SWI). Combining these different contrasts has the potential to improve the segmentation of the red nucleus. In the following, we will refer to these different contrasts as {\sl modalities}, keeping in mind that some are actually contrasts derived from a single sequence (iMag, R2* and QSM) while others are a separate sequence.

Multimodal medical image fusion has attracted great interest in recent years (please see~\cite{zhou2019review, hermessi2021multimodal} for reviews). 
Chartsias et al. \cite{chartsias2020disentangle} proposed a method that can improve segmentation accuracy of the modality of interest by learning to leverage the information from different modalities. 
The key idea is to learn a disentangled decomposition into anatomical and imaging factors.
This research demonstrates that using multimodal data through disentangled representations is an effective way to increase the robustness of the model and improve performance.
Guo et al. \cite{guo2019deep} proposed a model architecture to fuse cross-modality representation at feature learning, classifier and decision-making level, respectively. 
They conducted experiments using 4 imaging modalities and showed improvement compared to the baseline network using a single modality.
Chen et al. \cite{chen2019robust} proposed a multimodal learning model to decompose the disentanglement feature into modality-specific appearance and modality-invariant content, and share representation between different disentangled contents.
It can achieve good performance on brain tumor segmentation tasks.
All of the above studies demonstrated the benefits of multimodal learning, but have not effectively exploited the similarities and differences in information between modalities.
Moreover, these methods require to have all  modalities as input. However, it is frequent that one or more modalities are missing, noisy, or lack annotated data in real-world tasks \cite{rahate2022multimodal}. 

In this paper, we propose a multimodal fusion approach to segment the red nucleus based on various iron-sensitive modalities (iMag, QSM, R2*, SWI).
The underlying hypothesis of our work is that the anatomical structure of the target will be approximately constant, and that the cross-modal variation will mainly be in contrast. Such hypothesis is of course not universally true (there may be sequences that do not show or show only part of some anatomical structures), however we believe it holds in our application as well as in several others.
If the data can be decoupled into domain-invariant and domain-shift parts, then a large amount of computational resources can be saved by learning the domain-invariant part only once. Moreover, another advantage is that the approach does not require all modalities to be present for all subjects or to be annotated. Only the target modality needs to be systematically present and annotated.

\section{Methods} \label{sec:methods}
\begin{figure}[!hbtp]
\centering
\includegraphics[width=0.9\linewidth]{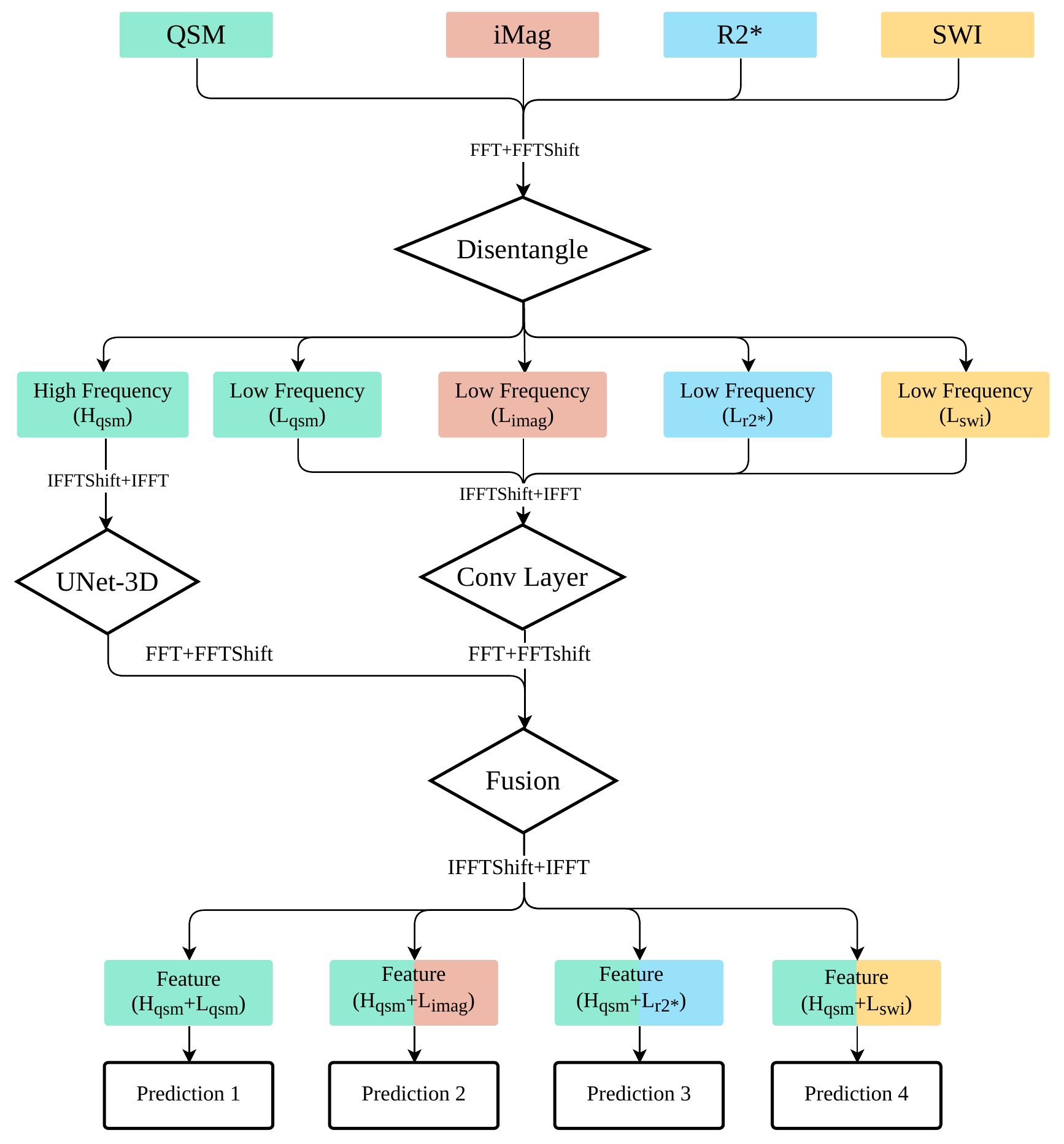}
\caption{Processing flow of the proposed method. The figure displays an example where the target modality is QSM for easy understanding but the principle would be similar with a different target.}
\label{fig:overall_flow}
\end{figure}

An overview of our approach can be seen in Figure~\ref{fig:overall_flow}.
To ease understanding, we show an example where the target is QSM data but the same concept can be used with another modality as a target. The inputs of the approach are the target modality (here QSM) and a list of other modalities that will bring low-frequency information (here iMag, R2* and SWI). The proposed model first disentangles each image into high- and low-frequency parts. The high-frequency (here $H_{qsm}$) of the target image is passed through a segmentation backbone (here a U-Net), and the low-frequency list (here $L_{qsm}, L_{imag}, L_{R2*}, L_{swi}$) is input to a shared weighted convolutional layer, and is then fused with the output of the segmentation backbone.

\subsection{Fourier domain disentangling}\label{sec:methods:disentangle}

Given $p$ modalities ${M_1, M_2, \ldots, M_p}$ with samples $i_{M_j} \in I_{M_j}$, where $I_{M_j} \subset \mathbb{R}^{H \times W}$ is a set of images, and $H$ and $W$ are the height and width. 
The disentangling operation is achieved by first transferring to Fourier space, and separating the high- and low-frequency components as follows:
\begin{equation}
\begin{aligned}
&{\scriptstyle \mathcal{F}({i_{M_j}}) = H^\theta_F({i_{M_j}}) + L^\theta_F({i_{M_j}})} \\
&{\scriptstyle L^\theta_F({i_{M_j}}) = \mathcal{F}({i_{M_j}})\left[\frac{H\times (1-\theta)}{2}: \frac{H\times(1+\theta)}{2},\frac{W\times(1-\theta)}{2}: \frac{W\times(1+\theta)}{2}\right]}\\
&{\scriptstyle H^\theta_F({i_{M_j}}) = \mathcal{F}({i_{M_j}}) - L_F({i_{M_j}}) }
\end{aligned}
\label{eq:fourier_disentangle}
\end{equation}
where $\mathcal{F}$ represents the Fourier transform, $L^\theta_F$ is the extraction of the low frequency part (in Fourier space, it is the center of the signal), $H^\theta_F$ is the extraction of the high frequency part (defined by subtracting the low-frequency part) and $\theta \in (0,1)$ is a parameter that controls the high-/low-frequency separation.

Our approach can get benefit from the low-frequency part of other modalities, even if not all modalities are present for all subjects. Actually, as will be shown in the experiments, even when only one subject has a given modality it can still be beneficial.
The low-frequency component, taken from a handful of subjects, can be seen as prior-knowledge.

\begin{figure}[!hbtp]
\centering
\includegraphics[width=0.9\linewidth]{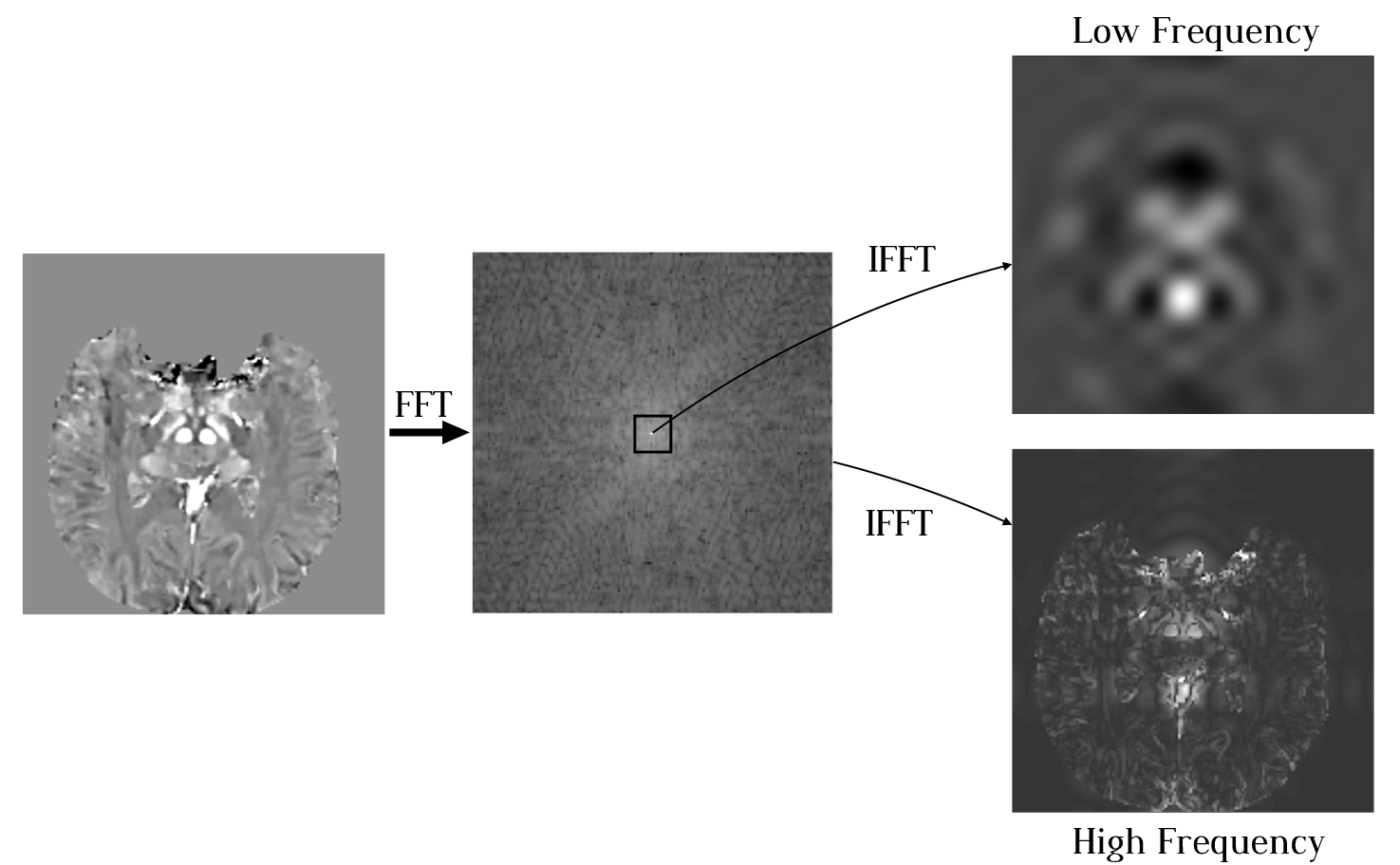}
\caption{Disentangled representation into high- and low-frequency parts of a QSM MRI. Note that we pad the low-frequency information to the original image size with 0 and then perform an inverse Fourier transform for visualization. Parameter $\theta$ (controlling the high-/low- frequency separation) is 0.1 in this figure.}
\label{fig:high_low_frequency_visualization}
\end{figure}

The visualization of the high- and low-frequency parts of the QSM can be seen in Figure~\ref{fig:high_low_frequency_visualization}.
The high-frequency part basically represents the brain structure and the low-frequency part the contrast of the modality.
The disentangled representation can help the model to learn brain structure and contrast information separately.

\subsection{Model architecture} \label{sec:method_model}
As previously mentioned, an image is first disentangled into high- and low-frequency parts. For the target modality $M_1$, we further use both high- and low-frequency parts $H^{1,\theta}_F$ and $L^{1,\theta}_F$ while, for other modalities $M_2,\ldots,M_p$, we only use the low-frequency part: $L^{2,\theta}_F, \ldots, L^{p,\theta}_F$. 
We then apply the inverse Fourier transform ($\mathcal{F}^{-1}$) to obtain high- and low-frequency parts in image space. The results are respectively denoted as $H^{1,\theta}_{F^{-1}}$ and $L^{j,\theta}_{{F}^{-1}}$ where $j\in\{1,\ldots,p\}$.
Then, $H^{1,\theta}_{{F}^{-1}}$ is fed to the backbone (a 3D-UNet \cite{cciccek20163d} in our experiments), the output being denoted as $O^{1,\theta}_{{F}^{-1}}$. 
The low-frequency list $L^{1,\theta}_{{F}^{-1}}, \ldots, L^{p,\theta}_{{F}^{-1}}$ is processed by a shared convolutional layer, the outputs being denoted as $O^{1,\theta}_{L_{F^{-1}}}, \ldots, O^{p,\theta}_{L_{F^{-1}}}$. $O^{1,\theta}_{L_{F^{-1}}}, \ldots, O^{p,\theta}_{L_{F^{-1}}}$ are then fused with $O^{1,\theta}_{H_{F^{-1}}}$ one by one. 
Each fusion operation sets the center values of $O^{1,\theta}_{H_{F^{-1}}}$ to $O^{j,\theta}_{L_{F^{-1}}}$:
$O_{H_{F^{-1}}}^{1,\theta} [\frac{H\times (1-\theta)}{2}: \frac{H\times(1+\theta)}{2},\frac{W\times(1-\theta)}{2}: \frac{W\times(1+\theta)}{2}] = O^{j,\theta}_{L_F^{-1}}$.
The high and low-frequency fusion operation thus produces $p$ outputs. They are further processed with convolution, activation function and dropout, producing $p$ predictions.
The final loss function is the sum of differences between the $p$ predictions and the ground truth.

\section{Experiments and Results} \label{sec:experiments}

\subsection{Dataset} \label{sec:experiments_dataset}
We used a dataset comprising 18 healthy subjects, 46 patients with early Parkinson’s disease (i.e. with a disease duration below 4 years), and 16 patients with prodromal parkinsonism (idiopathic rapid eye movement sleep behavior disorder-iRBD), recruited between May 2015 and January 2019 as part of the ICEBERG cohort.

\begin{figure}[!hbtp]
\centering
\includegraphics[width=\linewidth]{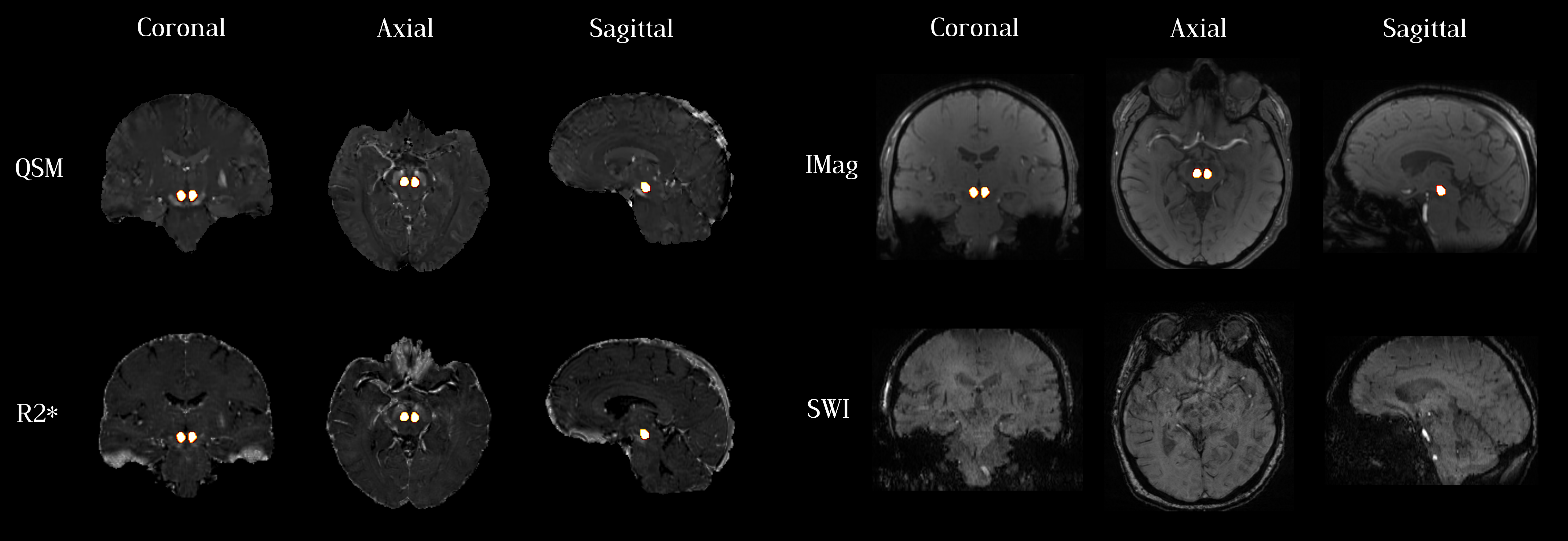}
\caption{Visualization of four modalities (QSM, R2*, iMag and SWI). The segmentation of the red nucleus is superimposed on the first three images. }
\label{fig:data_visualization}
\end{figure}

Each patient had a multi echo 3D GRE (12 echo times ranging from 4 ms to 37 ms) with a full brain coverage at an isotropic voxel resolution of 1~mm$^3$. 
The iMag contrast is the average over the echos of the magnitude volumes. 
The R2* is the exponential fit of the echo time decay, and the QSM is the fit of quantitative susceptibility.
We performed different experiments in which the target image to segment was either QSM, R2* or iMag. 
Additionally, we extracted low-frequency information from an SWI sample (without annotated ground truth) to enrich the low-frequency list.
Figure~\ref{fig:data_visualization} displays examples of these fours modalities.
The dataset was split at the participant-level into training, validation and test sets that contains 51, 13 and 16 participants separately. The split was stratified for diagnostic class.

\subsection{Implementation details}
Our code is developed based on the PyTorch framework \cite{paszke2019pytorch}. We used the open-source Python library TorchIO \cite{perez2021torchio} for reshaping images to the same size for each task and for min-max normalization. We used Adam \cite{kingma2014adam} as optimizer, Dice as loss function and a batch size of 1.

\subsection{Experiments}
In the following experiments, we compare the model performance for different modality combinations and with different amounts of training data. In all experiments, the hyperparameters were identical.

We report performance at the voxel level for a total of 7 metrics: Dice score, precision, recall, mean volume error rate (MVER), mean absolute volume error rate (MAVER), Pearson's $r$ and 95 percentile Hausdorff distance.

\subsubsection{Extracting prior-knowledge from different modalities}
We first experimented with varying the number of modalities bringing prior knowledge (low-frequency part only) in addition to the target modality (both high- and low-frequency parts). The prior knowledge may come from:
\begin{itemize}
    \item 0 modality. We simply use the target modality (decoupling the data into high and low-frequency). This is to evaluate the effect of disentanglement.
    \item 1 to 3 modality. For the modalities in our experiment, we mainly have two situations, similar contrast (QSM and R2*) or inverse contrast (iMag and SWI). These experiments can assess the effect of adding other modalities, including the contrast relationship and the number of modalities.
\end{itemize}

Since we are mainly interested in the performance in the low sample size regime, we only used 7.5\% of the training set for the experiment. Results can be seen in Table~\ref{tab:exp:modalities}.

The improvements obtained when adding prior knowledge modalities are not consistent: they depend on the metrics and the combination. However, there are cases where adding more modalities substantially improves a given metric as can for instance be seen for the 95 percentile Hausdorff.

\subsubsection{Varying the size of the training set}
We then evaluated the performance of our approach for different sizes of the training set and compared it to a baseline approach (UNet). Specifically, the models were trained with 7.5\% (4 subjects), 15\% (8 subjects), 30\% (16 subjects), 50\% (26 subjects) and 100\% (51 subjects) of the training set, while the validation and test sets were left unchanged.
The low-frequency part is only used as a prior knowledge and we actually used only one subject (from the training set) for the modalities used for low-frequency. We perform this experiment based on the best modalities combinations obtained for the Dice score in the previous results shown in Table~\ref{tab:exp:modalities} (iMag: iMag+R2*+SWI; QSM: QSM+SWI and R2*: iMag+QSM+R2*).
The results obtained with variable training set sizes are shown in Table~\ref{tab:exp:percentage}.
Our approach model outperforms the baseline in the very small training size regime (7.5\% data).
This demonstrates the potential of our algorithm with small training data sets, thus avoiding the expensive labeling.

\begin{table}[h!]
\centering
\caption{Results of different modality combination when using only 7.5\% of the training set (4 subjects). Results are reported as $\text{mean}\pm\text{SEM}$ where $\text{SEM}$ is the standard-error of the mean.}
\vspace{0.2cm}
\resizebox{1\linewidth}{!}{

\begin{tabular}{|l|llll|lllllll|} 
\hline
\multirow{2}{*}{Task} & \multicolumn{4}{c|}{Modality} & \multicolumn{7}{c|}{\textbf{Voxel level Performance}}                                                                                              \\ 
\cline{2-12}
                      & iMag & QSM & R2* & SWI        & Dice               & 95~Hausdorff        & Precision          & Recall             & MVER               & MAVER              & Pearson's r         \\ 
\hline
\multirow{8}{*}{iMag} & \checkmark    &     &     &            & 0.69±0.02          & 38.45±6.01          & 0.6±0.03           & 0.82±0.02          & 0.41±0.07          & 0.41±0.07          & 0.7±0.02            \\
                      & \checkmark    & \checkmark   &     &            & 0.58±0.03          & 54.74±3.91          & 0.45±0.03          & \textbf{0.84±0.02} & 0.98±0.14          & 0.98±0.14          & 0.61±0.02           \\
                      & \checkmark    &     & \checkmark   &            & 0.69±0.02          & 27.19±6.05          & 0.6±0.02           & 0.8±0.03           & 0.35±0.05          & 0.35±0.05          & 0.69±0.02           \\
                      & \checkmark    &     &     & \checkmark          & 0.7±0.02           & \textbf{20.32±5.57} & 0.63±0.02          & 0.8±0.03           & 0.28±0.05          & 0.29±0.05          & 0.7±0.02            \\
                      & \checkmark    & \checkmark   & \checkmark   &            & 0.65±0.03          & 46.29±5.53          & 0.56±0.03          & 0.79±0.03          & 0.47±0.08          & 0.47±0.08          & 0.66±0.02           \\
                      & \checkmark    & \checkmark   &     & \checkmark          & 0.67±0.02          & 30.77±6.14          & 0.59±0.03          & 0.79±0.02          & 0.38±0.07          & 0.39±0.07          & 0.68±0.02           \\
                      & \checkmark    &     & \checkmark   & \checkmark          & \textbf{0.72±0.02} & 25.17±6.40          & \textbf{0.65±0.02} & 0.81±0.03          & \textbf{0.27±0.05} & \textbf{0.27±0.05} & \textbf{0.72±0.02}  \\
                      & \checkmark    & \checkmark   & \checkmark   & \checkmark          & 0.69±0.03          & 24.52±6.49          & 0.61±0.03          & 0.81±0.03          & 0.39±0.09          & 0.39±0.08          & 0.7±0.03            \\ 
\hline
\multirow{8}{*}{QSM}  &      & \checkmark   &     &            & 0.79±0.02          & 29.54±7.27          & 0.72±0.02          & 0.9±0.02           & 0.26±0.03          & 0.26±0.03          & 0.8±0.01            \\
                      & \checkmark    & \checkmark   &     &            & 0.8±0.02           & 31.88±8.06          & \textbf{0.73±0.02} & 0.89±0.02          & 0.23±0.04          & 0.24±0.03          & 0.8±0.02            \\
                      &      & \checkmark   & \checkmark   &            & 0.79±0.02          & \textbf{11.45±5.76} & 0.73±0.02          & 0.86±0.03          & \textbf{0.19±0.05} & \textbf{0.23±0.04} & 0.79±0.02           \\
                      &      & \checkmark   &     & \checkmark          & \textbf{0.8±0.01}  & 12.48±6.06          & 0.71±0.02          & \textbf{0.92±0.01} & 0.31±0.03          & 0.31±0.03          & \textbf{0.8±0.01}   \\
                      & \checkmark    & \checkmark   & \checkmark   &            & 0.77±0.02          & 31.62±7.77          & 0.7±0.02           & 0.87±0.02          & 0.26±0.04          & 0.26±0.04          & 0.78±0.01           \\
                      & \checkmark    & \checkmark   &     & \checkmark          & 0.78±0.02          & 38.2±7.38           & 0.71±0.02          & 0.88±0.02          & 0.26±0.05          & 0.27±0.05          & 0.79±0.02           \\
                      &      & \checkmark   & \checkmark   & \checkmark          & 0.75±0.02          & 44.69±6.86          & 0.65±0.03          & 0.9±0.01           & 0.41±0.06          & 0.41±0.06          & 0.76±0.02           \\
                      & \checkmark    & \checkmark   & \checkmark   & \checkmark          & 0.76±0.02          & 37.06±6.84          & 0.68±0.02          & 0.86±0.02          & 0.28±0.04          & 0.28±0.04          & 0.76±0.02           \\ 
\hline
\multirow{8}{*}{R2*}  &      &     & \checkmark   &            & 0.64±0.03          & 44.11±6.94          & 0.56±0.03          & 0.77±0.04          & \textbf{0.39±0.06} & \textbf{0.39±0.06} & 0.65±0.03           \\
                      & \checkmark    &     & \checkmark   &            & 0.58±0.03          & 53.64±5.83          & 0.47±0.03          & 0.8±0.04           & 0.76±0.08          & 0.76±0.08          & 0.61±0.03           \\
                      &      & \checkmark   & \checkmark   &            & 0.62±0.03          & 53.5±4.86           & 0.5±0.02           & 0.81±0.03          & 0.65±0.07          & 0.65±0.07          & 0.64±0.02           \\
                      &      &     & \checkmark   & \checkmark          & 0.59±0.02          & 56.33±4.95          & 0.45±0.02          & 0.86±0.02          & 0.96±0.08          & 0.96±0.08          & 0.62±0.02           \\
                      & \checkmark    & \checkmark   & \checkmark   &            & \textbf{0.68±0.02} & \textbf{37.82±6.71} & \textbf{0.58±0.03} & 0.82±0.02          & 0.44±0.06          & 0.44±0.06          & \textbf{0.69±0.02}  \\
                      & \checkmark    &     & \checkmark   & \checkmark          & 0.58±0.04          & 59.51±3.15          & 0.47±0.03          & 0.76±0.05          & 0.65±0.07          & 0.65±0.07          & 0.59±0.04           \\
                      &      & \checkmark   & \checkmark   & \checkmark          & 0.64±0.03          & 46.64±4.48          & 0.55±0.03          & 0.79±0.03          & 0.49±0.07          & 0.49±0.07          & 0.65±0.03           \\
                      & \checkmark    & \checkmark   & \checkmark   & \checkmark          & 0.64±0.02          & 45.5±4.81           & 0.51±0.03          & \textbf{0.86±0.02} & 0.72±0.07          & 0.72±0.07          & 0.66±0.02           \\
\hline
\end{tabular}
}
\label{tab:exp:modalities}
\end{table}

\begin{table}[h!]
\centering
\caption{The baseline model (U-Net) and the proposed approach are compared when using different subsets of size n=4, 8, 16, 26 or 51 subjects, respectively corresponding to 7.5\%, 15\%, 30\%, 50\% and 100\% of the training set. Results are reported as $\text{mean}\pm\text{SEM}$ where $\text{SEM}$ is the standard-error of the mean.}
\vspace{0.2cm}
\resizebox{1\linewidth}{!}{
\begin{tabular}{|l|l|l|lllllll|} 
\hline
\multicolumn{1}{|l}{Model} & Task                  & n  & Dice               & 95~Hausdorff        & Precision          & Recall             & MVER               & MAVER              & Pearson's r         \\ 
\hline
\multirow{15}{*}{UNet}     & \multirow{5}{*}{IMag} & 4 & 0.37±0.02          & 66.09±3.10          & 0.23±0.01          & \textbf{0.93±0.02} & 3.16±0.17          & 3.16±0.17          & 0.46±0.02           \\
                           &                       & 8  & 0.33±0.02          & 59.02±1.68          & 0.2±0.01           & \textbf{0.98±0.01} & 4.19±0.24          & 4.19±0.24          & 0.44±0.01           \\
                           &                       & 16  & 0.76±0.02          & 37.59±5.33          & 0.66±0.03          & 0.92±0.01          & 0.42±0.05          & 0.42±0.05          & 0.78±0.02           \\
                           &                       & 26  & 0.83±0.01          & \textbf{3.88±1.77}  & \textbf{0.76±0.02} & 0.93±0.01          & \textbf{0.24±0.03} & \textbf{0.24±0.03} & 0.84±0.01           \\
                           &                       & 51 & 0.8±0.01           & 9.42±3.73           & 0.69±0.02          & 0.96±0.01          & 0.41±0.04          & 0.41±0.04          & 0.82±0.01           \\ 
\cline{2-10}
                           & \multirow{5}{*}{QSM}  & 4 & 0.46±0.03          & 70.04±1.52          & 0.31±0.03          & \textbf{0.93±0.03} & 2.27±0.24          & 2.27±0.24          & 0.53±0.03           \\
                           &                       & 8  & 0.84±0.01          & \textbf{5.66±3.91}  & 0.78±0.02          & 0.92±0.02          & \textbf{0.18±0.04} & \textbf{0.2±0.03}  & 0.85±0.01           \\
                           &                       & 16  & 0.77±0.01          & 19.97±6.63          & 0.64±0.02          & \textbf{0.99±0.01} & 0.56±0.05          & 0.56±0.05          & 0.79±0.01           \\
                           &                       & 26  & \textbf{0.88±0.01} & 1.0±0.0             & 0.8±0.01           & \textbf{0.97±0.01} & 0.21±0.02          & 0.21±0.02          & 0.88±0.01           \\
                           &                       & 51 & 0.88±0.01          & 1.0±0.0             & 0.81±0.02          & \textbf{0.97±0.01} & 0.2±0.02           & 0.2±0.02           & 0.89±0.01           \\ 
\cline{2-10}
                           & \multirow{5}{*}{R2*}  & 4 & 0.34±0.03          & 69.58±1.68          & 0.21±0.02          & \textbf{0.83±0.03} & 3.21±0.27          & 3.21±0.27          & 0.42±0.02           \\
                           &                       & 8  & \textbf{0.85±0.01} & \textbf{1.12±0.09}  & \textbf{0.78±0.02} & \textbf{0.92±0.01} & \textbf{0.19±0.03} & \textbf{0.19±0.03} & \textbf{0.85±0.01}  \\
                           &                       & 16  & \textbf{0.86±0.01} & 5.54±4.55           & \textbf{0.82±0.02} & 0.91±0.01          & \textbf{0.13±0.03} & \textbf{0.14±0.03} & \textbf{0.86±0.01}  \\
                           &                       & 26  & \textbf{0.85±0.01} & \textbf{1.0±0.0}    & 0.77±0.01          & 0.95±0.01          & \textbf{0.23±0.02} & \textbf{0.23±0.02} & \textbf{0.86±0.01}  \\
                           &                       & 51 & 0.86±0.02          & 1.0±0.0             & 0.79±0.04          & 0.95±0.02          & 0.21±0.07          & 0.21±0.07          & \textbf{0.87±0.02}  \\ 
\hline
\multirow{15}{*}{Ours}     & \multirow{5}{*}{IMag} & 4 & \textbf{0.72±0.02} & \textbf{25.17±6.40} & \textbf{0.65±0.02} & 0.81±0.03          & \textbf{0.27±0.05} & \textbf{0.27±0.05} & \textbf{0.72±0.02}  \\
                           &                       & 8  & \textbf{0.79±0.01} & \textbf{20.0±5.44}  & \textbf{0.69±0.02} & 0.91±0.01          & \textbf{0.33±0.03} & \textbf{0.33±0.03} & \textbf{0.79±0.01}  \\
                           &                       & 16  & \textbf{0.81±0.01} & \textbf{18.71±6.24} & \textbf{0.74±0.02} & 0.92±0.01          & \textbf{0.26±0.04} & \textbf{0.26±0.04} & \textbf{0.82±0.01}  \\
                           &                       & 26  & 0.83±0.01          & 7.45±3.75           & 0.75±0.02          & 0.93±0.01          & 0.25±0.04          & 0.25±0.04          & 0.84±0.01           \\
                           &                       & 51 & \textbf{0.85±0.01} & \textbf{7.46±4.41}  & \textbf{0.76±0.02} & 0.96±0.01          & \textbf{0.26±0.02} & \textbf{0.26±0.02} & \textbf{0.85±0.01}  \\ 
\cline{2-10}
                           & \multirow{5}{*}{QSM}  & 4 & \textbf{0.8±0.01}  & \textbf{12.48±6.06} & \textbf{0.71±0.02} & 0.92±0.01          & \textbf{0.31±0.03} & \textbf{0.31±0.03} & \textbf{0.8±0.01}   \\
                           &                       & 8  & \textbf{0.85±0.01} & 7.47±4.83           & 0.78±0.02          & \textbf{0.95±0.01} & 0.22±0.04          & 0.22±0.04          & \textbf{0.86±0.01}  \\
                           &                       & 16  & \textbf{0.86±0.01} & \textbf{1.0±0.0}    & \textbf{0.77±0.02} & 0.97±0.01          & \textbf{0.27±0.03} & \textbf{0.27±0.03} & \textbf{0.87±0.01}  \\
                           &                       & 26  & 0.87±0.01          & 1.0±0.0             & 0.8±0.01           & 0.96±0.01          & \textbf{0.2±0.02}  & \textbf{0.2±0.02}  & 0.88±0.01           \\
                           &                       & 51 & 0.88±0.01          & 1.0±0.0             & \textbf{0.82±0.01} & 0.96±0.01          & \textbf{0.18±0.02} & \textbf{0.18±0.02} & 0.89±0.01           \\ 
\cline{2-10}
                           & \multirow{5}{*}{R2*}  & 4 & \textbf{0.68±0.02} & \textbf{37.82±6.71} & \textbf{0.58±0.03} & 0.82±0.02          & \textbf{0.44±0.06} & \textbf{0.44±0.06} & \textbf{0.69±0.02}  \\
                           &                       & 8  & 0.78±0.01          & 3.12±1.04           & 0.69±0.02          & 0.89±0.02          & 0.3±0.05           & 0.32±0.04          & 0.78±0.01           \\
                           &                       & 16  & 0.84±0.01          & \textbf{1.06±0.035} & 0.76±0.01          & \textbf{0.96±0.01} & 0.28±0.03          & 0.28±0.03          & 0.85±0.01           \\
                           &                       & 26  & 0.84±0.01          & 1.03±0.03           & \textbf{0.77±0.02} & 0.95±0.01          & 0.24±0.03          & 0.24±0.03          & 0.85±0.01           \\
                           &                       & 51 & 0.86±0.01          & 1.0±0.0             & \textbf{0.79±0.02} & \textbf{0.95±0.01} & \textbf{0.2±0.03}  & \textbf{0.2±0.03}  & 0.86±0.01           \\
\hline
\end{tabular}
}
\label{tab:exp:percentage}
\end{table}

\section{Conclusion} \label{sec:Conclusion}
In this paper, we proposed a new approach for segmentation of the red nucleus from brain MRI data. The approach can leverage the low-frequency information from different contrasts (or modalities). Learning high-frequency and low-frequency information separately allows the model to focus on brain structure throughout the learning process while contrast information is fused in the final stage.

Our experiments show that the proposed approach leads to substantial improvements when dealing with very small training sets. 
On the other hand, the addition of other modalities led to mixed results, providing improvements only part of the time. Future work shall focus on understanding how and why our Fourier disentanglement process can lead to such massive improvements in performance over the baseline in the very small training set regime. 

Other future improvements would be to use convolution theory to do multiplication in the Fourier domain to improve computational efficiency, exploring other methods for representation disentanglement and applying the framework to other datasets. 

\section{Compliance with ethical standards}
The institutional ethical standard committee approved the study (CPP Paris VI/RCB: 2014-A00725-42). 
All participants gave written informed consent.

\section{Acknowledgments}\label{sec:acknowledgments}
The research leading to these results has received funding from the French government under management of Agence Nationale de la Recherche as part of the "Investissements d'avenir" program, reference ANR-19-P3IA-0001 (PRAIRIE 3IA Institute) and reference ANR-10-IAIHU-06 (Agence Nationale de la Recherche-10-IA Institut Hospitalo-Universitaire-6).  
The ICEBERG study is supported by the European Research Council (ERC) under grant agreement No. 678304, the European Union’s Horizon 2020 research and innovation program under grant agreement No. 826421 (TVB-Cloud), Agence Nationale de la Recherche (ANR) under grant agreements ANR-10-IAIHU-06 (IHU ICM), ANR-11-INBS-0006, and ANR-19-JPW2-000 (JPND E-DADS), association France Parkinson (PRECISE-PD project), the Fondation d’Entreprise EDF, Biogen Inc., Fondation Thérèse and René Planiol, Fondation Saint Michel. It received unrestricted support for Research on Parkinson’s disease from Energipole (M. Mallart), M. Villain and the Société Française de Médecine Esthétique (M. Legrand).
Guanghui Fu is supported by the Chinese Government Scholarship provided by China Scholarship Council (CSC). Lydia Chougar is supported by a Poste d'accueil Inria/AP-HP.

\bibliographystyle{IEEEbib}
\bibliography{refs}
\end{document}